%% file: main.tex
\RequirePackage{lineno}
\documentclass[aps,prd,twocolumn,groupedaddress,amsmath,amssymb]{revtex4-2}

\usepackage{amssymb}
\usepackage{overpic}

\usepackage{graphicx}
\usepackage{subfigure}
\usepackage{dcolumn}
\usepackage{bm}
\usepackage{epsfig}
\usepackage{xcolor}
\usepackage{enumerate}
\usepackage{booktabs}
\usepackage{multirow}
\usepackage{hyperref}
\usepackage{lipsum}
\hypersetup{hidelinks,
	colorlinks=true,
	allcolors=blue,
	pdfstartview=Fit,
	breaklinks=true}

\begin{document}

\newcommand{\jpsi}{J/\psi}
\newcommand{\pip}{\pi^+}
\newcommand{\pin}{\pi^-}
\newcommand{\pio}{\pi^0}
\newcommand{\g}{\gamma}
\newcommand{\gev}{GeV/c$^2$}
\newcommand{\mev}{MeV/c$^2$}
\newcommand{\ar}{\rightarrow}
\newcommand{\ks}{K_S^{0}}
\newcommand{\etap}{\eta^\prime}
\let\oldequation\equation
\let\oldendequation\endequation
\renewenvironment{equation}{\linenomathNonumbers\oldequation}{\oldendequation\endlinenomath}
\title{\boldmath Measurement of the Branching Fraction of \boldmath{$\psi(2S) \to \gamma \pi^0$} }

\author{
\begin{center}
\input{authorlist_2024-03-31}
\end{center}
}


\begin{abstract}
Based on $(2712.4\pm14.1)\times10^{6}~\psi(2S)$ events, 7.9 fb$^{-1}$ $\psi(3773)$ data, and 0.8 fb$^{-1}$ off-resonance data samples collected with the BESIII detector, we measure the branching fraction of $\psi(2S)\rightarrow\gamma\pi^{0}$ and $e^{+}e^{-}\rightarrow\gamma\pi^{0}$ form factor at momentum transfers $Q^{2}\sim13$ GeV$^{2}$. The $e^{+}e^{-}\rightarrow\gamma\pi^{0}$ cross section is fitted with considering the interference between the $\psi(2S)$ and continuum amplitudes and two solutions are found, ${\cal B}=3.74\times10^{-7}$ with $\phi=3.93$ rad and ${\cal B}=7.87\times10^{-7}$ with $\phi=2.08$ rad. Here, ${\cal B}$ is the branching fraction of $\psi(2S)\rightarrow\gamma\pi^{0}$ and $\phi$ is the relative phase angle between the $\psi(2S)$ and continuum amplitudes. Due to insufficient off-resonance data, the branching fraction ${\cal B}(\psi(2S)\rightarrow\gamma\pi^{0})$ is determined to be in the range $[2.7, 9.7]\times10^{-7}$ within one standard deviation of the contour region.
\end{abstract}

\maketitle

\section{Introduction}
Within the framework of Quantum Chromodynamics (QCD), vector charmonium radiative decays are believed to proceed mainly via short-distance $c\bar{c}$ annihilations into two gluons with a photon emitted by charm quarks~\cite{gg}. However, when considering vector charmonium radiative decay to a light pseudoscalar, other phenomenological mechanisms become important, such as the mixing of $\eta_{c}-\eta^{(\prime)}$~\cite{eta1,eta2}, final state radiation by light quarks~\cite{finalrad}, and the vector meson dominance (VMD) model~\cite{finalrad,vmd1}. Measurements of these charmonium radiative decays provide tests for various theoretical predictions. 

The decay rate of $\psi(2S)\rightarrow\gamma\pi^{0}$ is expected to be smaller than those of $\psi(2S)\rightarrow\gamma\eta$ and $\gamma\eta^{\prime}$ due to the suppressed gluon coupling to isovector currents~\cite{finalrad}.  BESIII has measured the branching fractions of decays $\psi(2S)\rightarrow\gamma\pi^{0}$, $\gamma\eta$, and $\gamma\eta^{\prime}$ and reported ${\cal B}(\psi(2S)\rightarrow\gamma\pi^0)=(9.5\pm1.6\pm0.5)\times10^{-7}$~\cite{Gpi0}. The branching fractions of $\psi(2S)\rightarrow\gamma\eta$ and $\gamma\eta^{\prime}$ are well understood using the VMD model associated with the mixing of $\eta_{c}-\eta^{(\prime)}$. However,  the branching fraction of $\psi(2S)\rightarrow\gamma\pi^{0}$ is one order of magnitude larger than the predicted range of $(0.66\sim1.15)\times10^{-7}$ from the VMD model~\cite{zhaoq}. 

 This discrepancy may arise from the irreducible background contribution from the continuum amplitude of $e^{+}e^{-}\rightarrow\gamma^{*}\rightarrow\gamma\pi^{0}$. 
 In the recent BESIII measurement~\cite{Gpi0}, the contribution from the continuum amplitude was not considered due to insufficient data samples at off-resonance energy points, and the continuum itself and its interference with the $\psi(2S)$ decay amplitude were neglected. With increased statistics of both off-resonance and resonance data, the interference effect can now be taken into account, and the $e^{+}e^{-}\rightarrow\gamma^{*}\rightarrow\gamma\pi^{0}$ form factor can be measured through the continuum process $e^{+}e^{-}\rightarrow\gamma^{*}\rightarrow\gamma\pi^{0}$. This allows for a more reliable determination of branching fraction of $\psi(2S)\rightarrow\gamma\pi^{0}$.

The cross section for the process $e^{+}e^{-}\rightarrow\gamma^{*}\rightarrow\gamma\pi^{0}$ at center-of-mass (c.m.) energy $\sqrt{s}$ is given by    
    \begin{equation}
		\sigma(s)=\frac{2\pi^{2}\alpha^{3}}{3}|F_{\pi^0}(s)|^{2},
		\label{eq:pi0}
	\end{equation}
	where $F_{\pi^0}$ is the $\pi^{0}$ transition form factor and $\alpha$ is fine-structure constant. The form factor $F_{\pi^0}$ is calculated using perturbative quantum chromodynamics (pQCD) in the asymptotic limit~\cite{ff1,ff2},
		\begin{equation}
	     -Q^2F_{\pi^{0}}(Q^2)=\sqrt{2}f_{\pi^{0}}\left(1-\frac{5}{3}\frac{\alpha_{s}(Q^2)}{\pi}\right),~Q^2\gg m_{\pi^0}^2,
	\end{equation}
	where $Q^2=-q^{2}$, $q$ is the four-momentum of
the virtual photon, $f_{\pi^{0}}$ is the $\pi^{0}$ decay constant, $m_{\pi^0}$ is the mass of $\pi^{0}$~\cite{pdg}, and $\alpha_{s}$ is the strong coupling. This measurement provides a critical test of pQCD predictions~\cite{ff1,ff2}.
	 
    In this paper, we measure the branching fraction of $\psi(2S)\rightarrow\gamma\pi^{0}$ based on 2.7 billion $\psi(2S)$ events, 7.9 fb$^{-1}$ $\psi(3773)$ data, and 0.8 fb$^{-1}$ off-resonance data samples collected at $\sqrt{s}=$ 3.650 and 3.682 GeV with the BESIII detector operating at the BEPCII~\cite{BESIII:2024lks,psipp,psipp2024}.

\section{BESIII detector and MC simulation}
The BESIII detector~\cite{Ablikim:2009aa} records symmetric $e^+e^-$ collisions 
provided by the BEPCII storage ring~\cite{Yu:2016cof}
in the c.m. energy range from 1.84 to  4.95~GeV,
with a peak luminosity of $1.1 \times 10^{33}\;\text{cm}^{-2}\text{s}^{-1}$ 
achieved at $\sqrt{s} = 3.773\;\text{GeV}$. 
BESIII has collected large data samples in this energy region~\cite{Ablikim:2019hff,EcmsMea,EventFilter}. The cylindrical core of the BESIII detector covers 93\% of the full solid angle and consists of a helium-based
 multilayer drift chamber~(MDC), a plastic scintillator time-of-flight
system~(TOF), and a CsI(Tl) electromagnetic calorimeter~(EMC),
which are all enclosed in a superconducting solenoidal magnet
providing a 1.0~T magnetic field.
The solenoid is supported by an
octagonal flux-return yoke with resistive plate counter muon
identification modules interleaved with steel. 
The charged-particle momentum resolution at $1~{\rm GeV}/c$ is
$0.5\%$, and the 
${\rm d}E/{\rm d}x$
resolution is $6\%$ for electrons
from Bhabha scattering. The EMC measures photon energies with a
resolution of $2.5\%$ ($5\%$) at $1$~GeV in the barrel (end cap)
region. The time resolution in the TOF barrel region is 68~ps, while
that in the end cap region was 110~ps.  The end cap TOF
system was upgraded in 2015 using multigap resistive plate chamber
technology, providing a time resolution of
60~ps, which benefits ~84$\%$ of the data
used in this analysis~\cite{Tof1,Tof2,Tof3}.

Monte Carlo (MC) simulated data samples produced with a {\sc
geant4}-based~\cite{geant4}  software package, which
includes the geometric description of the BESIII detector and the
detector response, are used to determine detection efficiencies
and to estimate backgrounds. The simulation models the beam
energy spread and initial state radiation (ISR) in the $e^+e^-$
annihilations with the generator {\sc
kkmc}~\cite{kkmc1,kkmc2}. Inclusive MC simulations are used to estimate backgrounds including the production of the
$\psi(2S)$ resonance, the production of $D\bar{D}$
pairs (including quantum coherence for the neutral $D$ channels),
the non-$D\bar{D}$ decays of the $\psi(3770)$, the ISR production , and
the continuum processes incorporated in {\sc
kkmc}~\cite{kkmc1,kkmc2}. All particle decays are modeled with {\sc
evtgen}~\cite{evtgen1,evtgen2} using branching fractions 
either taken from the
Particle Data Group~\cite{pdg}, when available,
or otherwise estimated with {\sc lundcharm}~\cite{lund1,lund2}. Final state radiation~(FSR)
from charged final state particles is incorporated using the {\sc
photos} package~\cite{photos}. The signal MC for $e^{+}e^{-}\rightarrow\gamma\pi^{0}$ is generated to determine the efficiency using the helicity amplitude (HELAMP) model with the {\sc
evtgen}~\cite{evtgen1,evtgen2}.

\section{Event selection} 
\label{sec:fit}
The $\pi^{0}$ is reconstructed with two photons so there are three photons in the final state and no good charged tracks.

Good charged tracks detected in the MDC are required to be within a polar angle ($\theta$) range of $|\rm{cos\theta}|<0.93$, where $\theta$ is defined with respect to the $z$-axis, which is the symmetry axis of the MDC. The distance of closest approach to the interaction point (IP) must be less than 10\,cm along the $z$-axis, and less than 1\,cm in the transverse plane.

Photon candidates are identified using isolated showers in the EMC. The photons are required to be only in barrel EMC ($\lvert \cos\theta\rvert\textless$0.8) to suppress the backgrounds from the quantum electrodynamics (QED) processes and the deposited energy of each photon in EMC should be more than 25 MeV. The number of photons is required to be exactly 3, which can suppress most backgrounds that involve more than 3 photons in the final state. A four-constraint (4C) kinematic fit with the constraints provided by four-momentum conservation is imposed on the three photons, and $\chi_{\rm 4C}^2$ is required to be less than 40. The photon with the highest energy is regarded as the radiative photon and other two photons with lower energy are regarded as the $\pi^{0}$ candidates. To further suppress backgrounds from the QED process  $e^+e^-\rightarrow\gamma\gamma\gamma$, the cosine of the helicity angle of $\gamma$ is required to be less than 0.7. The helicity angle is defined as the angle between the momentum of the $\gamma$ from $\pi^{0}$ with larger deposited energy in the $\pi^{0}$ rest frame and the momentum of $\pi^{0}$ in the $e^{+}e^{-}$ c.m. frame. 

\begin{figure}[htbp]
		\centering
		\includegraphics[width=0.95\linewidth]{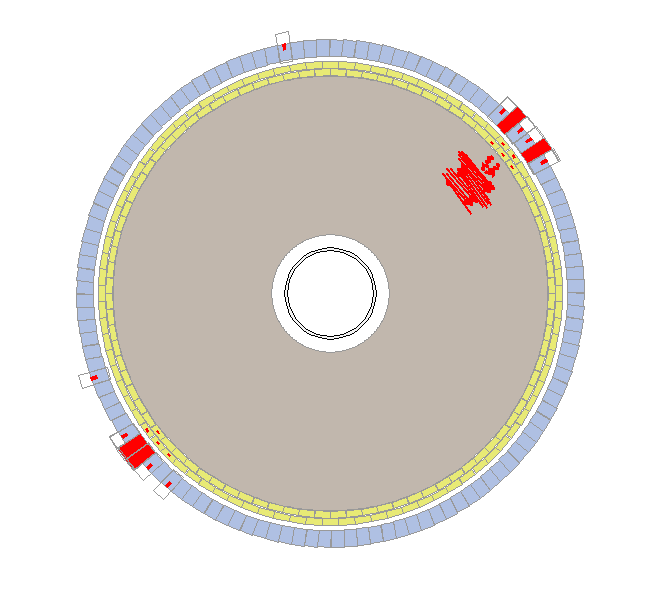}
		\caption{The view of a $\gamma$-conversion event in the MDC and EMC in $x-y$ plane.}
        \label{fig:mdchit}
	\end{figure}

\begin{figure*}[htbp]
	    \centering
		\subfigure{}{
				\centering
				\includegraphics[width=0.48\linewidth]{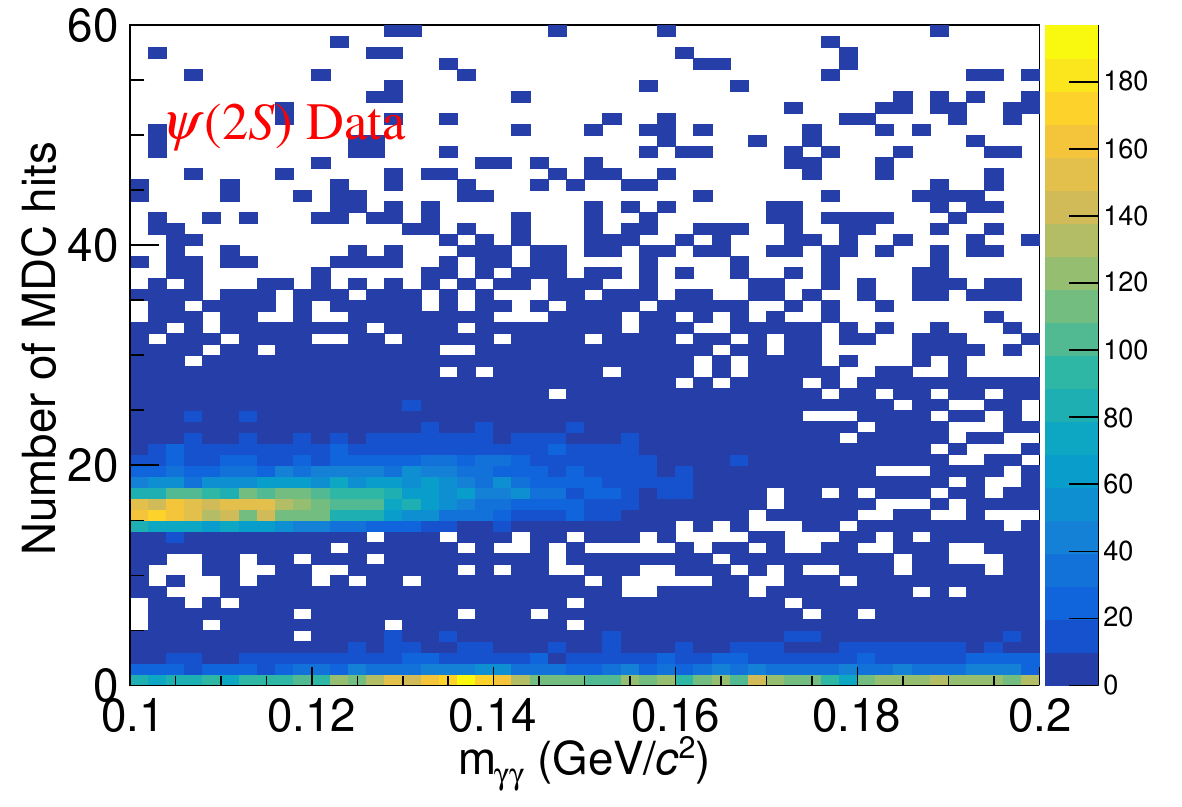}
		}
		\subfigure{}{
				\includegraphics[width=0.48\linewidth]{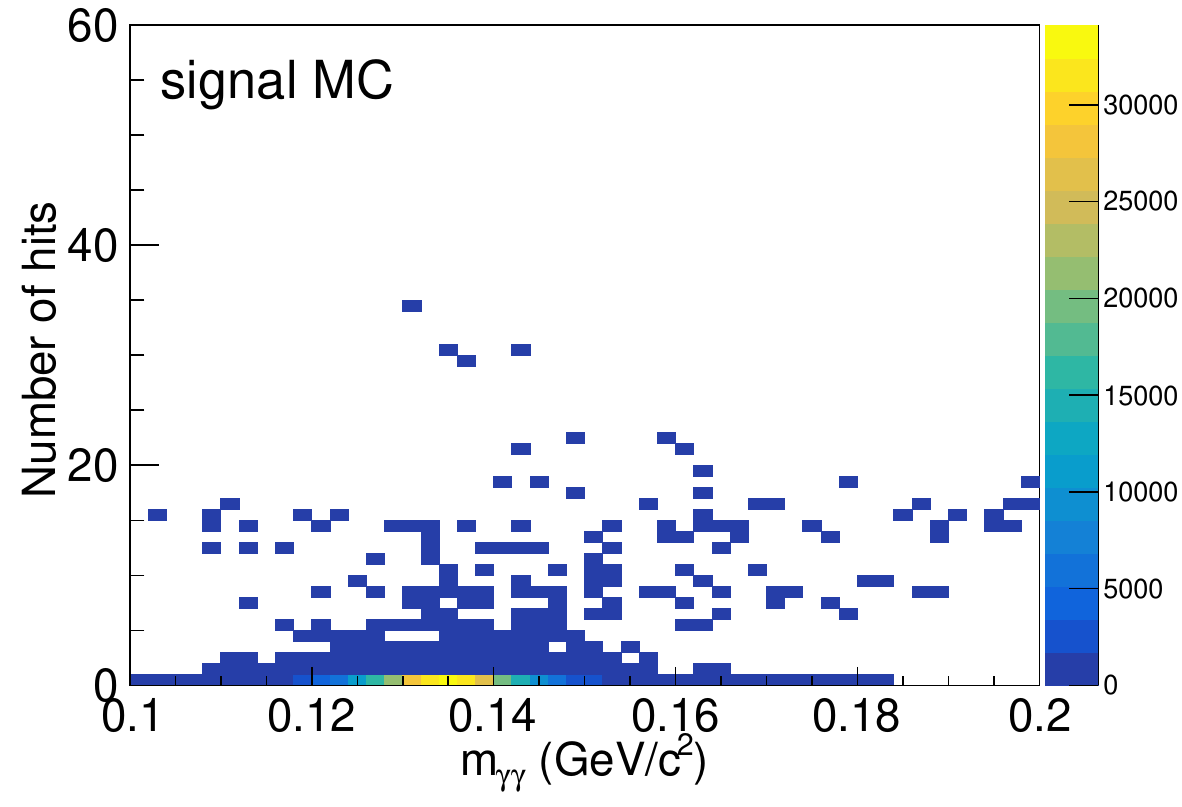}
		}
		\caption{Two-dimensional distributions of the number of hits versus $m_{\gamma\gamma}$ in $\psi(2S)$ data (left panel) and MC-simulated signal process (right panel).}
        \label{fig:hit2D}
	\end{figure*}

The background is studied with inclusive $\psi(2S)$ MC samples and QED processes $e^{+}e^{-}\rightarrow \gamma\gamma(\gamma_{\rm ISR})$ generated at each energy point. The dominant backgrounds are the $e^{+}e^{-}\rightarrow \gamma\gamma$ events where one photon converts into an $e^{+}e^{-}$ pair, and if the track finding algorithm fails, electrons and positrons can be misidentified as two photons. An example of a $\gamma$-conversion event in the $x-y$ plane is displayed in Fig.~\ref{fig:mdchit}. Figure~\ref{fig:hit2D} shows the two-dimensional distribution of the number of hits versus the $\gamma\gamma$ invariant mass ($m_{\gamma\gamma}$) in the $\psi(2S)$ data and MC simulation. To suppress these background events, the number of hits in the MDC in the region between the two radial lines connecting the IP and the two shower positions in the EMC is required to be less than 8~\cite{BESIII:2010tfr}. Most of the $\gamma$-conversion events from the $e^{+}e^{-}\rightarrow \gamma\gamma$ process can be rejected. After applying the above requirements, the remaining peaking backgrounds come from $\psi(2S)\rightarrow\gamma\chi_{cJ}$ decays with $\chi_{cJ}\rightarrow\pi^{0}\pi^{0}$, where $J$=0 or 2.
The contribution from these peaking backgrounds is estimated precisely using the well-measured branching fractions~\cite{pdg}. The selection efficiencies are 0.017\% and 0.045\% for $\psi(2S)\rightarrow\gamma\chi_{c0}$ and $\psi(2S)\rightarrow\gamma\chi_{c2}$, repectively. The expected numbers are 125.1$\pm$12.5 for $\psi(2S)\rightarrow\gamma\chi_{c0}$ and 85.6$\pm$10.0 for $\psi(2S)\rightarrow\gamma\chi_{c2}$. The other remaining smooth backgrounds are from QED processes ($e^{+}e^{-}\rightarrow \gamma\gamma\gamma(n\gamma),\ n\ge0$). Due to the inefficiency of the reconstruction algorithms of photons, background events with more than three photon candidates still survive after requiring exactly three photon candidates.


To extract the number of signal events, an unbinned maximum likelihood fit is performed on the $m_{\gamma\gamma}$ spectrum for each dataset. When calculating $m_{\gamma\gamma}$, the measured momentum of $\gamma$ is used instead of the momentum modified by 4C kinematic fit to avoid possible bias. 

For the fit to $m_{\gamma\gamma}$ at $\sqrt{s}=3.686$ GeV, the probability density function (P.D.F.) consists of three components: the signal, the peaking background from $\psi(2S)\rightarrow\gamma\chi_{cJ}$ decays with $\chi_{cJ}\rightarrow\pi^{0}\pi^{0}$, and the smooth background from QED processes ($e^{+}e^{-}\rightarrow \gamma\gamma\gamma(n\gamma),\ n\ge0$). The signal is modeled using a MC-simulated shape convolved with a Gaussian function to account for the difference in mass resolution between data and MC simulation. The mean and standard deviation of the Gaussian function are free in the fit. The peaking background is described by the MC-simulated shape, and the number of peaking background events is fixed to the expected value from MC studies. The smooth background is described by a third-order Chebychev polynomial function with shape parameters obtained from MC-simulated QED processes. 

For the fits at $\sqrt{s}=$ 3.650, 3.682, and 3.773 GeV, the P.D.F only consists of a signal and a smooth background from MC-simulated QED processes. The parameters of the Gaussian function in the signal shape at $\sqrt{s}=$ 3.650 and 3.682 GeV are constrained to the values obtained from the fit at $\sqrt{s}=$ 3.686 GeV due to limited statistics at these energies. 

\begin{figure*}[htbp]
    \centering
		\subfigure{
		\begin{overpic}[width=0.48\linewidth,height=6cm]{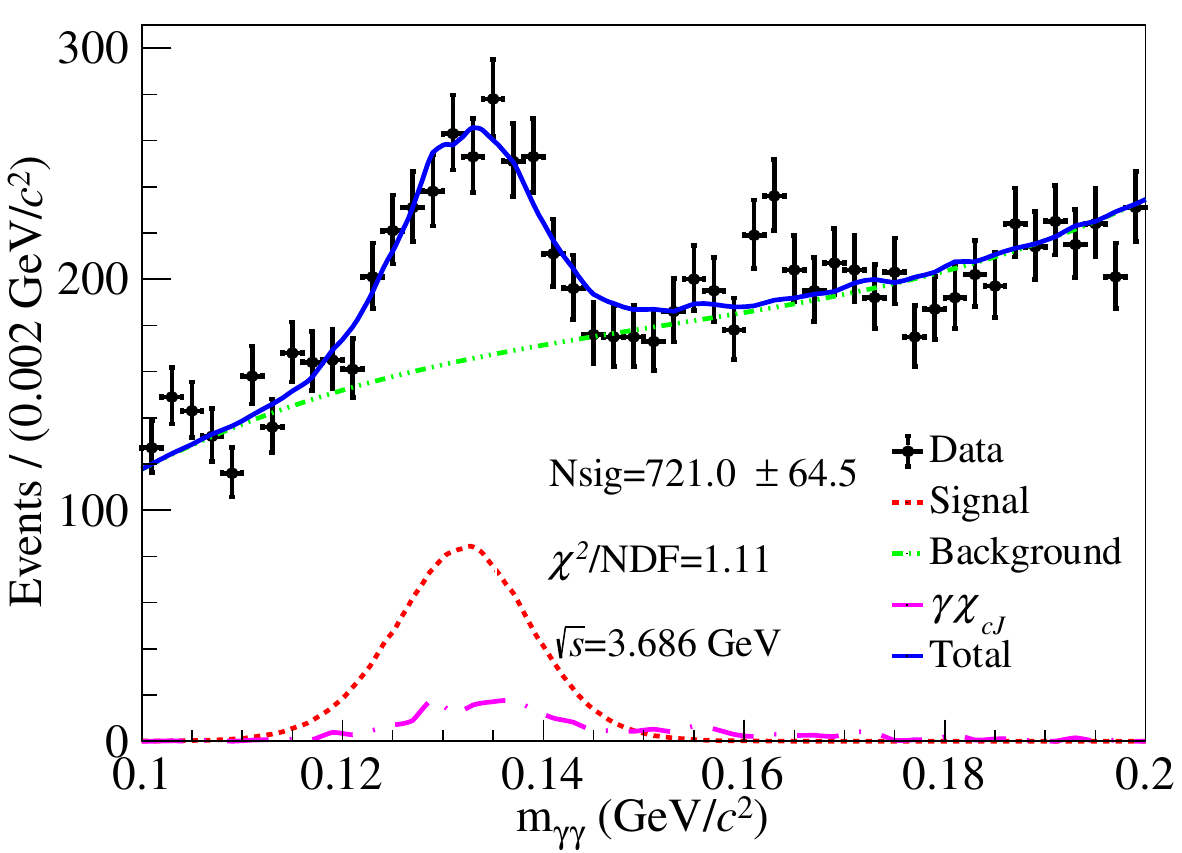}
		\put(15,60){\large (a)}
		\end{overpic}
		}
		\subfigure{
		\begin{overpic}[width=0.48\linewidth,height=6cm]{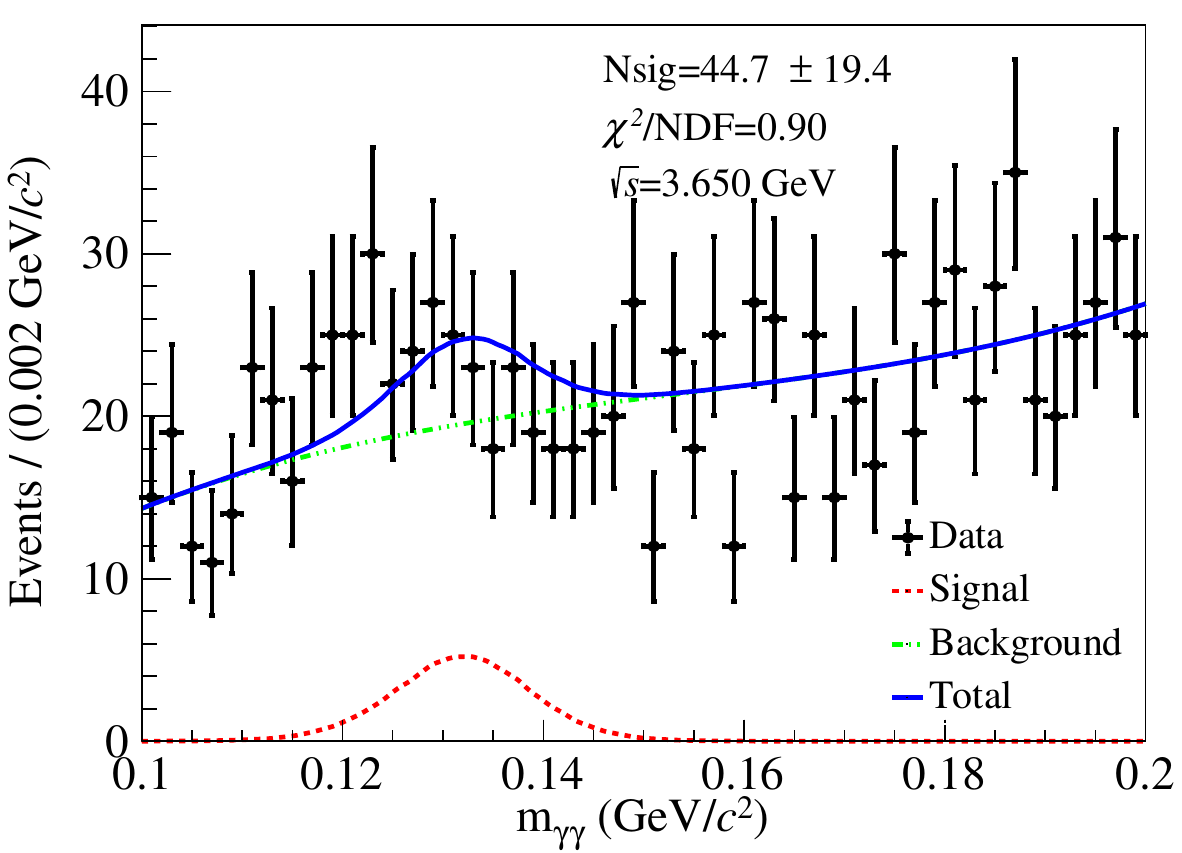}
		\put(15,60){\large (b)}
		\end{overpic}
		}
		\subfigure{
		\begin{overpic}[width=0.48\linewidth,height=6cm]{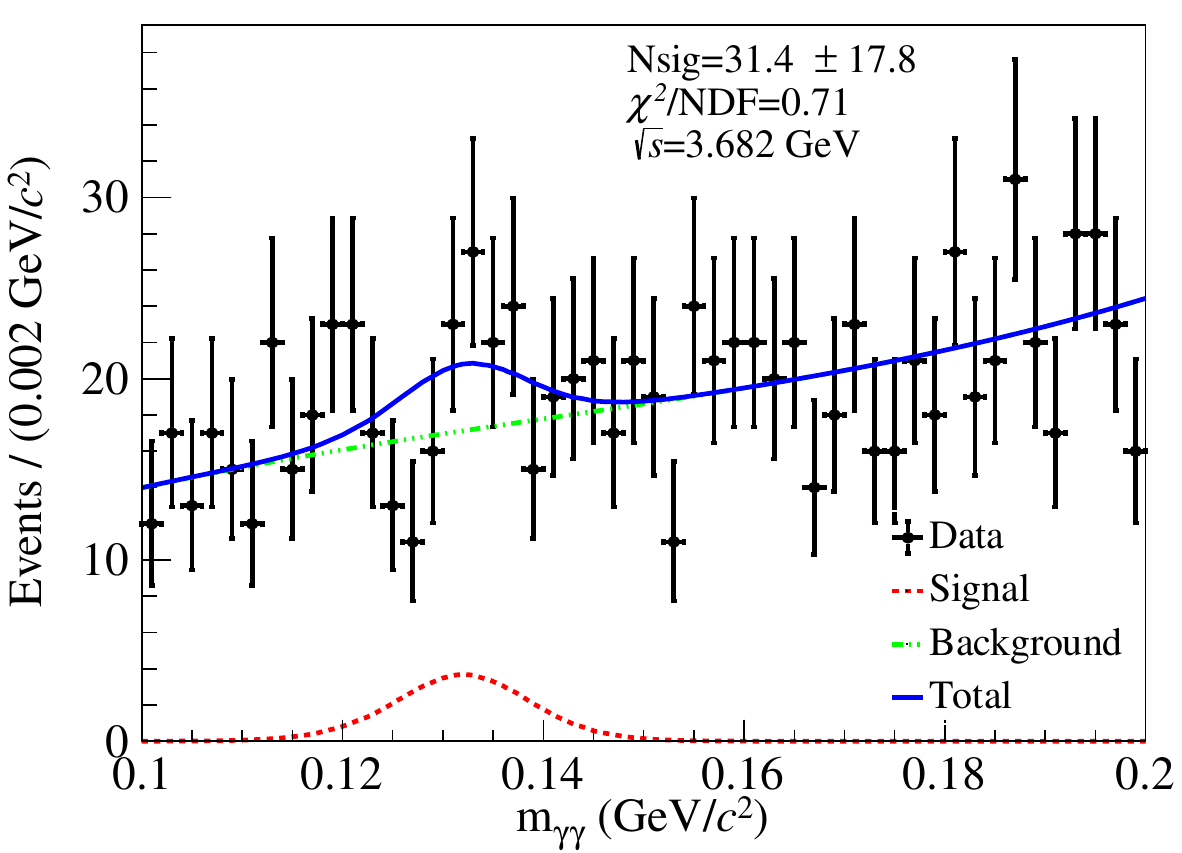}
		\put(15,60){\large (c)}
		\end{overpic}
		}
		\subfigure{
		\begin{overpic}[width=0.48\linewidth,height=6cm]{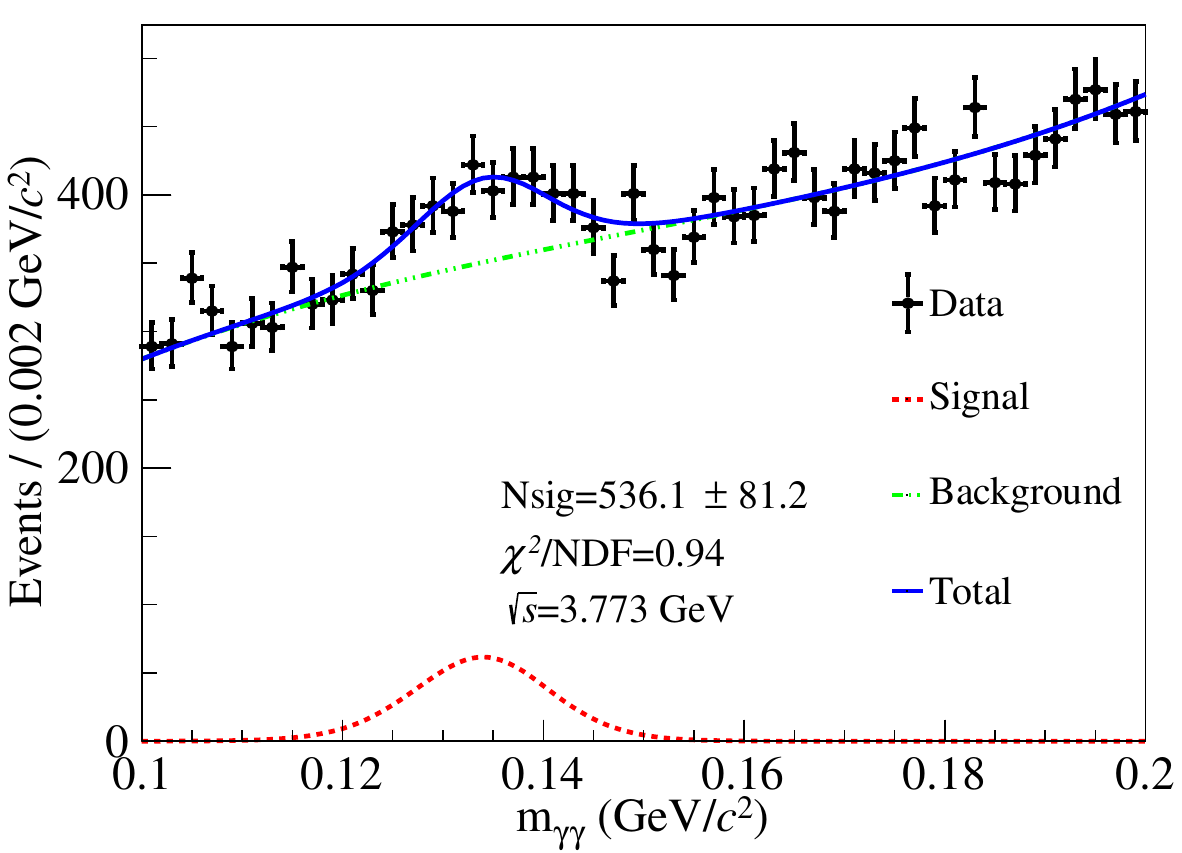}
		\put(15,60){\large (d)}
		\end{overpic}
		}
		\caption{The fit to the $m_{\gamma\gamma}$ at the c.m. energy points of (a) $\psi(2S)$; (b) $\sqrt{s}=3.650$ GeV; (c) $\sqrt{s}=3.682$ GeV; (d) $\sqrt{s}=3.773$ GeV. Dots with error bars represent data, blue solid curves represent fit results, red dashed curves represent signal MC simulation, green dashed curves represent QED background contributions, and the pink dash-dotted curve represents peaking background from $\psi(2S)\rightarrow\gamma\chi_{cJ}$ decays.}
        \label{figure:fit}
	\end{figure*}

The fit results at each c.m. energy are shown in Fig.~\ref{figure:fit}. The statistical significances are 12.2$\sigma$, 2.4$\sigma$, 1.8$\sigma$, and 6.2$\sigma$ in comparison to an alternative fit without including the signal shape at $\sqrt{s}=$ 3.686, 3.650, 3.682, and 3.773 GeV, respectively. The signal yields and the observed cross sections for all data samples are summarized in Table~\ref{tab:sig}.
\begin{table*}[htbp]
\caption{The signal yields and the observed cross sections at each c.m. energy, where $N^{\rm sig}$ is the signal yield, $L$ is the integrated luminosity, $\epsilon$ is selection efficiency, and $\sigma^{\rm obs}$ is the observed cross section.}
    \label{tab:sig}
    \centering
    \begin{tabular}{ccccc}
    \hline\hline
      $\sqrt{s}$ (GeV) & $L$ (pb$^{-1}$)& $N^{\rm sig}$ &  $\epsilon$ (\%) &$\sigma^{\rm obs}$ (pb)\\ \hline
      3.650  &  401  & $44.7\pm19.4$     & 29.00 &  $0.38\pm0.17$ \\
      3.682  &  392  & $  31.4\pm 17.8$ &  29.86 &  $0.27\pm0.15$\\
     3.686 & 3877&$721.0\pm 64.5$ & 30.62  & $0.61\pm0.05$\\
     3.773    & 7909   & $536.1\pm81.2$  & 30.84 &$0.22\pm0.03$\\ \hline\hline
    \end{tabular}
\end{table*}

\section{Branching Fraction Measurement}
\label{section:br}
At each c.m. energy, the observed cross section is calculated with \\
	\begin{equation}
		\sigma^{\rm obs}=\frac{N^{\rm sig}}{L\cdot\epsilon\cdot{{\cal B}_{\pi^{0}\rightarrow{\gamma}{\gamma}}}},
	\end{equation}
	where $N^{\rm sig}$ is the number of signal events obtained from the fit to the $m_{\gamma\gamma}$ distribution as described in Sec.~\ref{sec:fit}, $L$ is the integrated luminosity, $\epsilon$ is the selection efficiency, and ${\cal B}_{\pi^{0}\rightarrow{\gamma}{\gamma}}$ is the branching fraction of $\pi^{0}\rightarrow\gamma\gamma$ taken from PDG~\cite{pdg}. The efficiency is estimated by an iterative weighting method~\cite{isr} and results are shown in Table~\ref{tab:sig}. 
	
	The cross section of $e^+e^-\rightarrow\gamma\pi^{0}$ near the vicinity of the $\psi(2S)$ peak is a coherent sum of the continuum amplitude and the resonance amplitude of $\psi(2S)$ which is given by
	\begin{widetext}
\begin{equation}
		\sigma_{\rm tot}(s)=\frac{4\pi\alpha^{2}s}{3}\left|\frac{f\cdot e^{i\phi}}{s^{\frac{3}{2}}}\sqrt{\frac{\alpha\pi}{2}}+\frac{3}{\sqrt{s}\alpha}\frac{\sqrt{\Gamma_{\psi(2S)}\Gamma_{ee}{\cal B}(\psi(2S)\rightarrow\gamma\pi^{0})}}{s-M^{2}+iM\Gamma_{\psi(2S)}}\right|^{2}, 
	\end{equation}
 \end{widetext}
where the first term is the contribution of $e^{+}e^{-}\rightarrow\gamma^{*}\rightarrow\gamma\pi^{0}$ with the $F_{\pi^{0}}(s)$ in Eq.(\ref{eq:pi0}) parameterized as $\frac{f}{s}$, where $f$ is a free parameter in the fit. The second term is the relativistic Breit-Wigner function of $\psi(2S)$, $\phi$ is the relative phase between the $\psi(2S)$ and continuum amplitudes, $\Gamma_{\psi(2S)}$, $\Gamma_{ee}$, and $M$ are the total width, $e^{+}e^{-}$ partial width, and mass of $\psi(2S)$, respectively, and are fixed to their PDG values~\cite{pdg}.
As $\psi(2S)$ is a narrow resonance, ISR and beam energy spread are taken into account in the fit and the final expression of the cross section is given by~\cite{Alexander:1988em}
	\begin{equation}
	\footnotesize
		\begin{split}
		& \sigma_{\rm exp}=\int_{0}^{x_{m}}dx\int_{\sqrt{s}-50\Delta}^{\sqrt{s}+50\Delta}d\sqrt{s'}F(x,s')\sigma_{\rm tot}(s'(1-x))G(\sqrt{s},\sqrt{s'}),\\
		&G(\sqrt{s},\sqrt{s'})=\frac{1}{\sqrt{2\pi}\Delta}e^{-\frac{(\sqrt{s}-\sqrt{s'})^{2}}{2\Delta^{2}}},
		\end{split}
	\end{equation}
	where $\Delta$= 1.3 MeV is the c.m. energy spread and $F(x,s)$ is the radiator function, which is calculated with an accuracy of 0.1\%  and approximated as~\cite{radfunc}
	\begin{widetext}
	\begin{eqnarray}
	    F(x,s)=x^{\beta-1}\beta\cdot(1+\delta)-\beta(1-\frac{x}{2})+\frac{1}{8}\beta^{2}\left[4\cdot(2-x)\cdot\ln{\frac{1}{x}}-\frac{1+3(1-x)^{2}}{x}\ln(1-x)-6+x\right],
	    \end{eqnarray}
	\end{widetext}
where $\delta=\frac{3}{4}\beta+\frac{\alpha}{\pi}(\frac{\pi^{2}}{3}-\frac{1}{2})+\beta^{2}(\frac{9}{32}-\frac{\pi^{2}}{12})$ and $\beta=\frac{2\alpha}{\pi}(\ln{\frac{s}{m_e^{2}}}-1)$.
A least $\chi^2$ fit to the measured cross sections versus $\sqrt{s}$ is performed with three free parameters $f$, $\phi$, and ${\cal B}(\psi(2S)\rightarrow\gamma\pi^{0})$. The $\chi^2$ is defined as
\begin{equation}
\text{{\fontsize{7.5}{10}\selectfont $\chi^{2}=\sum\limits_{i}\frac{\left(\sigma_{i}^{\rm obs}-\sigma^{\rm exp}(\sqrt{s}_{i})\right)^{2}}{(\Delta\sigma_{i}^{\rm obs})^2+\left(\Delta\sqrt{s}_{i}\cdot\frac{d\sigma^{\rm exp}(\sqrt{s}_{i})}{d\sqrt{s}}+\frac{1}{2}(\Delta\sqrt{s}_{i})^2\cdot\frac{d^{2}\sigma^{\rm exp}(\sqrt{s}_{i})}{d\sqrt{s}^{2}}\right)^2}$,}}
\end{equation}
where $i$ denotes the $i^{\rm th}$ energy point and $\Delta\sigma^{\rm obs}$ is the combination of statistical and uncorrelated systematic uncertainties. The second term in denominator estimates the uncertainty in the cross section due to the uncertainty of the c.m. energy~\cite{stat}.
	
\begin{figure}[htbp]
		\centering
		\subfigure{\label{subfig:solution1}}{
				\includegraphics[width=0.9\linewidth]{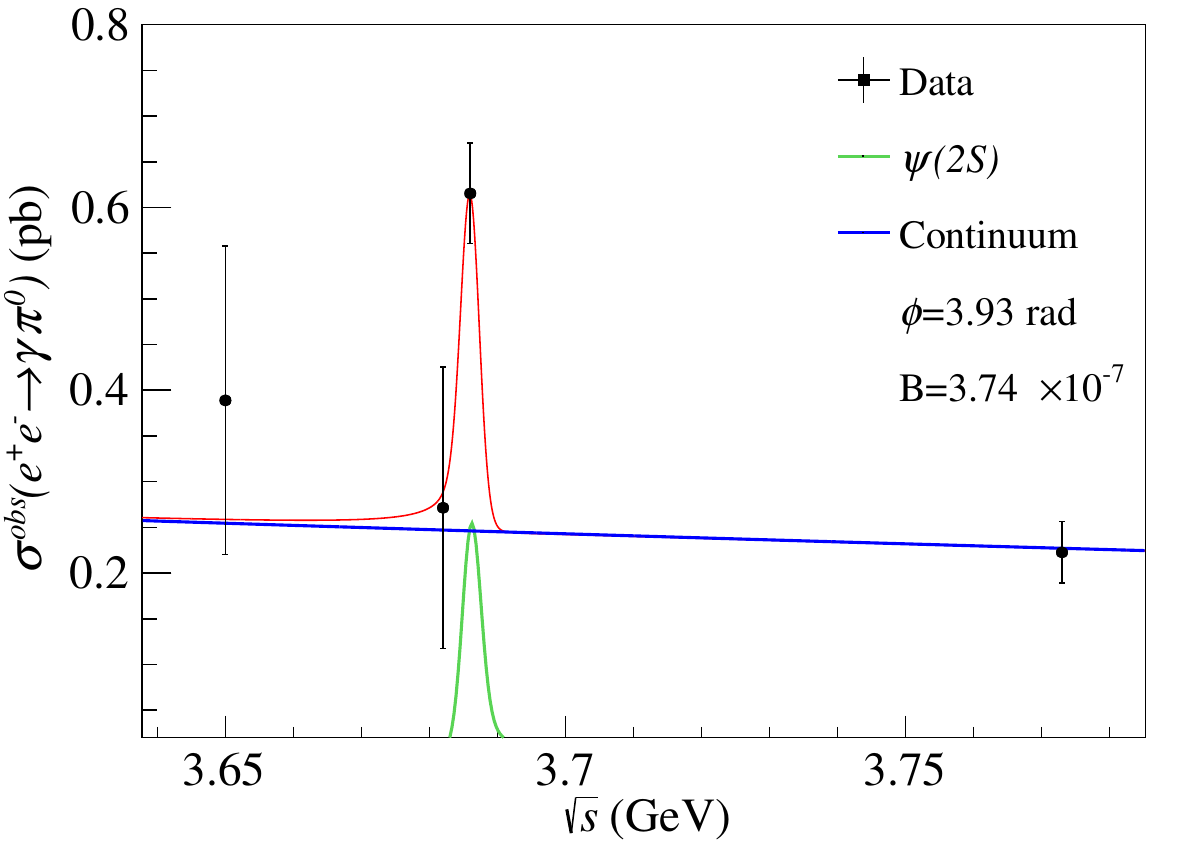}
		}
		\subfigure{\label{subfig:solution2}}{
				\includegraphics[width=0.9\linewidth]{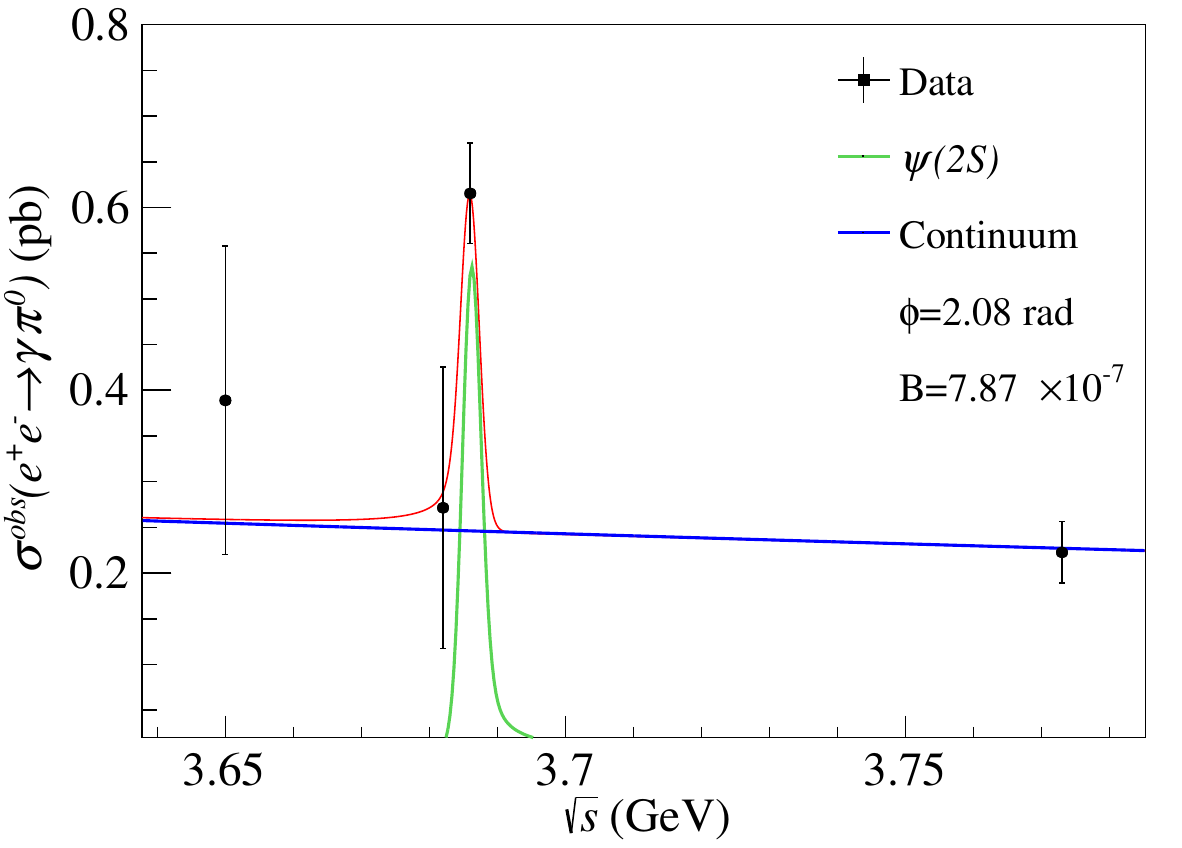}
		}
		\caption{Two optimal solutions of the fit to the observed cross section of $e^{+}e^{-}\rightarrow\gamma\pi^{0}$ in the vicinity of $\psi(2S)$ peak. Dots with error bars represent data, red solid lines denote fit results, green lines represent the contribution of $\psi(2S)$, blue lines represent the continuum contribution. Top panal is for solution a: ${\cal B}(\psi(2S)\rightarrow\gamma\pi^{0})=3.74\times10^{-7},\ \phi=3.93$ rad, and bottom panel is for solution b: ${\cal B}(\psi(2S)\rightarrow\gamma\pi^{0})=7.87\times10^{-7},\ \phi=2.08$ rad.}
        \label{fig:solution}
	\end{figure}

\begin{figure}[htbp]
		\centering
		\includegraphics[width=0.95\linewidth]{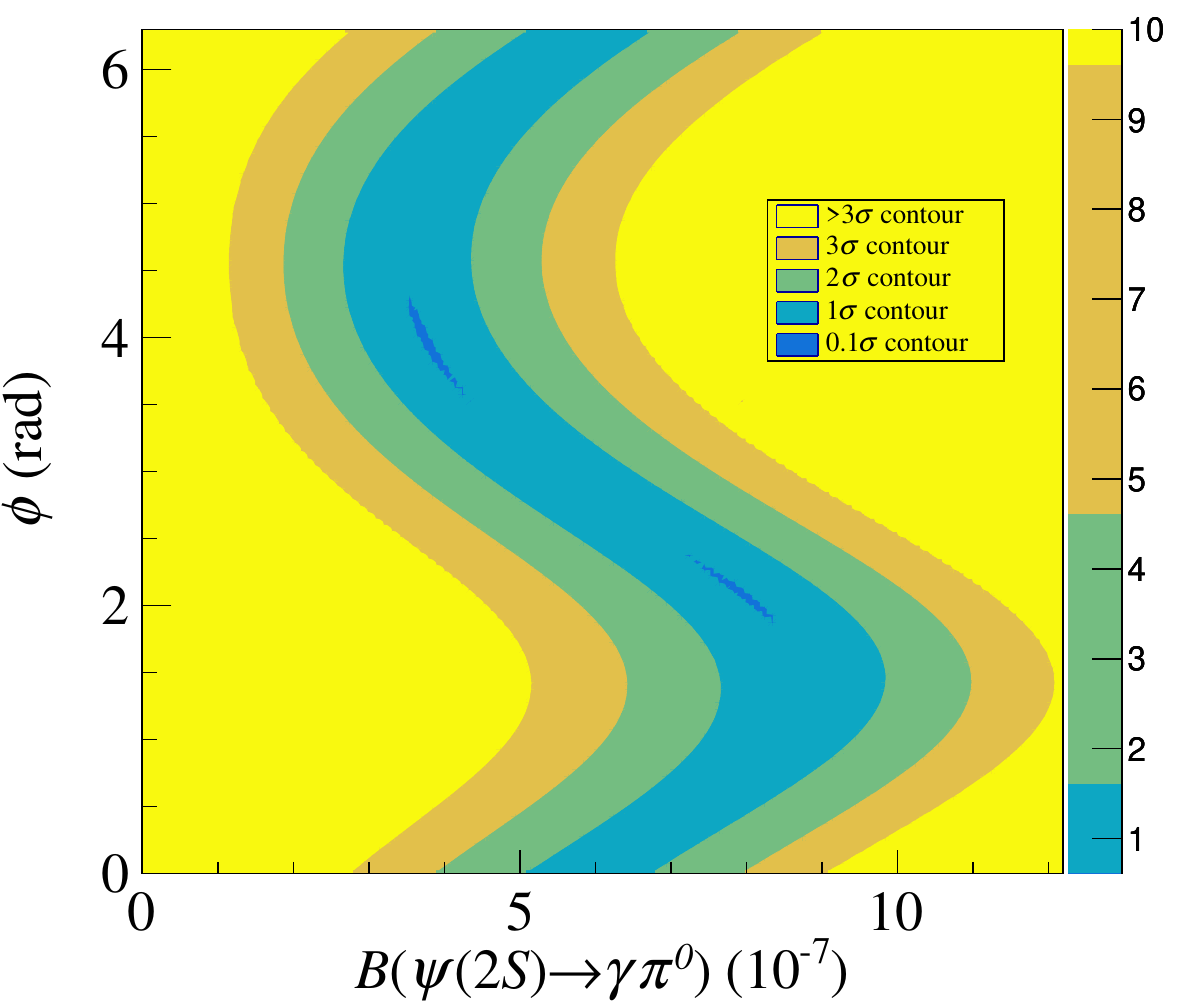}
		\caption{The two-dimensional scanning result of $\phi$ and ${\cal B}(\psi(2S)\rightarrow\gamma\pi^{0})$. The color represents the value of $\chi^{2}$.}       
		\label{fig:scan}
		\end{figure}

Two optimal solutions are found as the consequence of the form of the fit formula as shown in Fig.~\ref{fig:solution}~\cite{Zhu:2011ha}. Table~\ref{tab:solution} shows detailed information about the two solutions. The uncertainties of the two solutions are not presented because the phase between $\psi(2S)$ and continuum amplitudes can not be determined due to the low statistics of the off-resonance data. We perform a two-dimensional parameters scan of $\phi$ and ${\cal B}(\psi(2S)\rightarrow\gamma\pi^{0})$ to present the confidence interval of the fit result. Figure~\ref{fig:scan} shows the scanning result, where the value in each bin is the minimum $\chi^{2}$ of the fit with $\phi$ and ${\cal B}(\psi(2S)\rightarrow\gamma\pi^{0})$ fixed and $f$ free. The steps of $\phi$ and ${\cal B}(\psi(2S)\rightarrow\gamma\pi^{0})$ are 0.05 rad and 0.1$\times10^{-7}$, respectively. The contours of $\pm$0.5$\sigma$, $\pm$1$\sigma$, $\pm$2$\sigma$, and $\pm$3$\sigma$ are corresponding to the value of $\chi^{2}_{min}+0.25$, $\chi^{2}_{min}+1$, $\chi^{2}_{min}+4$, and $\chi^{2}_{min}+9$, respectively. The 1$\sigma$ contour is defined as the standard deviation of parameters approximately corresponding to the probability 39.3\%. The 1$\sigma$ contour range of ${\cal B}(\psi(2S)\rightarrow\gamma\pi^{0})$ is [2.7, 9.7]$\times10^{-7}$. The fit result of the form factor $f=\left|Q^{2}F_{\pi^{0}}(Q^{2})\right|$ is ($0.215\pm0.015$) GeV in this energy region, where the uncertainty is the combination of statistical and uncorrelated systematic uncertainties. 

	\begin{table}[htbp]
 	\centering
 	\caption{The information of two solutions.}
 	\label{tab:solution}
 		\begin{tabular}{ccccc}
 			\hline\hline
 			&$\chi^2$ & ${\cal B}(\psi(2S)\rightarrow\gamma\pi^{0})$     & $\phi$ (rad)&$f$ (GeV) \\
 			\hline
 			solution a&0.611  & $3.74\times 10^{-7}$ &3.93& $0.215$\\
 			solution b&0.611 & $7.87\times 10^{-7}$&2.08& $0.215$\\
 			\hline\hline
 		\end{tabular}
 \end{table}
\section{Systematic uncertainty}
 Systematic uncertainties for the observed cross section include uncertainties from many sources, such as the integrated luminosity measurements, photon reconstruction efficiency, $\pi^0$ branching fraction, 4C kinematic fit, requirements on MDC hits, requirements on the number of photons, signal shape, and background shape. Each of these uncertainties is discussed in detail below:
\subsection{Integrated luminosity}
    The uncertainties of the integrated luminosity measurement are studied using $e^+e^-\to \ell^+\ell^-(\gamma\gamma)~(\ell= e,\mu)$ events with an unctertainty less than 1\%~\cite{BESIII:2024lks,psipp,psipp2024}. We conservatively take 1\% as the corresponding systematic uncertainty.
    \subsection{Photon detection efficiency}
    The uncertainty for photon detection is 0.5\% per photon within barrel EMC region by studying $J/\psi\rightarrow\rho^{0}\pi^{0}$ process~\cite{BESIII:2016hfo} and total uncertainty is 1.5\%.
    \subsection{\boldmath{$\pi^0$} branching fraction}
    The uncertainty of the branching fraction of $\pi^0\rightarrow{\gamma\gamma}$ in PDG~\cite{pdg} is considered as systematic uncertainty which is 0.03\%.
    \subsection{4C kinematic fit}
    The uncertainty of kinematic fit is studied by using a control sample $e^{+}e^{-}\rightarrow\gamma\gamma\gamma$ and the difference of the efficiency between the MC simulation and data is taken as the systematic uncertainty which is determined to be 1.0\%~\cite{Gpi0}.
    \subsection{Requirement on MDC hits}
    Based on the control sample $\psi(2S)\rightarrow\gamma\chi_{c2}$ with $\chi_{c2}\rightarrow\gamma\gamma$ which has same final state as the signal process. The corresponding region used to count the number of hits in the MDC in this control sample is larger than that in the $\psi(2S)\to\gamma\pi^{0}$ decay due to the smaller Lorentz boost of the $\gamma\gamma$ system, and as a consequence, more hits from noise in the MDC will be counted in the control sample. To minimize this effect, the MDC hits are normalized according to area by assuming the noise is distributed uniformly over the MDC. The difference of the efficiency between data and MC simulation is determined to be 1.0\%~\cite{Gpi0}.
    \subsection{Requirement on the number of photons}
    The efficiency of the requirement on the number of photons is also studied with the control sample $\psi(2S)\rightarrow\gamma\chi_{c2}$ with $\chi_{c2}\rightarrow\gamma\gamma$. The data-MC difference is regarded as the systematic uncertainty which is 3.1\%~\cite{Gpi0}. 
    \subsection{Signal shape}
    In the extraction of the number of signal events, the signal is described by a MC-simulated shape convolved with a Gaussian function with the mean and standard deviation as free parameters. However, for the fit on samples at $\sqrt{s}=$ 3.650 and 3.682 GeV, the parameters of the Gaussian function are fixed due to low statistics at these energies. We change the values of the parameters of Gaussion function by one standard deviation from the fit at $\sqrt{s}=$ 3.686 GeV and the differences are estimated as the systematic uncertainty. The results of uncertainties are 3.1\% and 2.4\% at $\sqrt{s}=$ 3.650 and 3.682 GeV, respectively.
    \subsection{Background shape}
    In the fit to $\psi(2S)$ sample, the size of peaking backgrounds contribution from $\psi(2S)\rightarrow\gamma\chi_{cJ}$ decays with $\chi_{cJ}\rightarrow\pi^{0}\pi^{0}$ are fixed. We change the expected number of $\gamma\chi_{cJ}$ by $\pm1$ standard deviation in the fit and the largest difference is taken as the systematic uncertainty, which is 2.4\%. 
    
    To study the uncertainty associated with smooth background shape, the background shape is changed from a third-order Chebychev function to a fourth-order Chebychev function in the fit and the difference is considered as the systematic uncertainty. The uncertainties are 3.1\%, 1.4\%, 0.2\%, and 1.6\% at $\sqrt{s}=$ 3.650, 3.682, 3.686, and 3.773 GeV, respectively.
\subsection{Summary of the Systematic Uncertainty}
All of the sources of the systematic uncertainties are listed in Table~\ref{tab:uncertainty}. The systematic uncertainties of the cross section measurements are divided into correlated and uncorrelated uncertainties. The correlated uncertainties are common to all energies including branching fraction of $\pi^{0}\rightarrow\gamma\gamma$, the kinematic fit, the requirement on the number of photons, and the requirement on the number of MDC hits. The uncorrelated uncertainties are different at each energy including the signal shape, the background shape. The total uncertainty is calculated by 
\begin{equation}
    \sigma=\sqrt{\sum\limits_{i}\sigma_{i}^{2}},
\end{equation}
where $i$ denotes the $i^{\rm th}$ source of uncertainties assuming there is no correlation among them.
 	\begin{table*}[htp]
	\centering
	\caption{Summary of systematic uncertainties (\%) in the measurement of the $e^{+}e^{-}\rightarrow\gamma\pi^{0}$ cross section.}
 	\label{tab:uncertainty}
		\begin{tabular}{ccccc}
		\hline\hline
        Sources & 3.650 GeV&3.682 GeV&3.686 GeV&3.773 GeV\\
		\hline

		Photon reconstruction efficiency  & 1.5& 1.5& 1.5& 1.5\\
		${\cal B}(\pi^{0}\rightarrow\gamma\gamma)$  & 0.03& 0.03& 0.03& 0.03\\
		Requirement on $N_{\gamma}$  & 3.1& 3.1& 3.1& 3.1 \\
		Kinematic fit  & 2.0& 2.0& 2.0& 2.0 \\
		Number of MDC hits & 1.0& 1.0& 1.0& 1.0 \\
		Integrated luminosity&   1.0&   1.0&   1.0&   1.0\\
		Signal shape& 3.1&2.4&-& -\\
		Peaking background&-&-&2.4&-\\
		Smooth background shape&3.1&1.4&0.2&1.6\\
		\hline
		Total & 6.1 & 5.1&4.9&4.5\\
		\hline\hline
		\end{tabular}
	\end{table*}
 \section{summary}
Based on $\psi(2S)$, $\psi(3770)$, and off-resonance data samples collected at 3.650 and 3.682 GeV, we measure the branching fraction of $\psi(2S)\rightarrow\gamma\pi^{0}$ and the form factor of $e^{+}e^{-}\rightarrow\gamma\pi^{0}$ at $-Q^{2}\sim13~\rm{GeV}^{2}$. In the fit with the interference between $\psi(2S)$ and continuum amplitudes, two optimal solutions are found: ${\cal B}(\psi(2S)\rightarrow\gamma\pi^0)=3.74\times10^{-7},\ \phi=3.93$ rad and ${\cal B}(\psi(2S)\rightarrow\gamma\pi^0)=7.87\times10^{-7},\ \phi=2.08$ rad. The allowed range of ${\cal B}(\psi(2S)\rightarrow\gamma\pi^{0})$ is [2.7, 9.7]$\times10^{-7}$ with 1$\sigma$ contour of $\chi^2$ value. Comparing to the previous BESIII measurement of ${\cal B}(\psi(2S)\rightarrow\gamma\pi^0)=(9.5\pm1.7)\times10^{-7}$~\cite{Gpi0}, the branching fraction tends to be smaller due to interference effects, consistent with the VMD prediction~\cite{zhaoq} within uncertainties. The contribution from continuum process $e^{+}e^{-}\rightarrow\gamma^{*}\rightarrow\gamma\pi^{0}$ is important and non-negligible in the measurement of ${\cal B}(\psi(2S)\rightarrow\gamma\pi^0)$. This measurement supports the VMD model's description of the vector charmonium radiative decays to light pseudoscalars.

In the framework of pQCD, the form factor is calculated as
	\begin{equation}
	\begin{split}
	     \left|Q^2F_{\pi^{0}}(Q^2)\right|&=\sqrt{2}f_{\pi^{0}}\left(1-\frac{5}{3}\frac{\alpha_{s}(Q^2)}{\pi}\right)\\
	     &=\sqrt{2}f_{\pi^{0}}\approx0.185\ {\rm GeV}\ {\rm at}\ Q^{2}\rightarrow\infty,\\
	     \end{split}
	\end{equation}
In this measurement,  $f=\left|Q^{2}F_{\pi^{0}}(Q^{2})\right|=(0.215\pm0.015\pm0.004)$ GeV at $-Q^{2}\sim13~\rm{GeV}^{2}$, where the first uncertainty is the combination of statistical and uncorrelated systematic uncertainties, the second uncertainty is correlated systematic uncertainty, and the total uncertainty is 0.016 GeV. This result is slightly larger than the pQCD value but agrees with it within 1.9$\sigma$ deviation. 

\begin{figure}[htbp]
		\centering
		\includegraphics[width=0.95\linewidth]{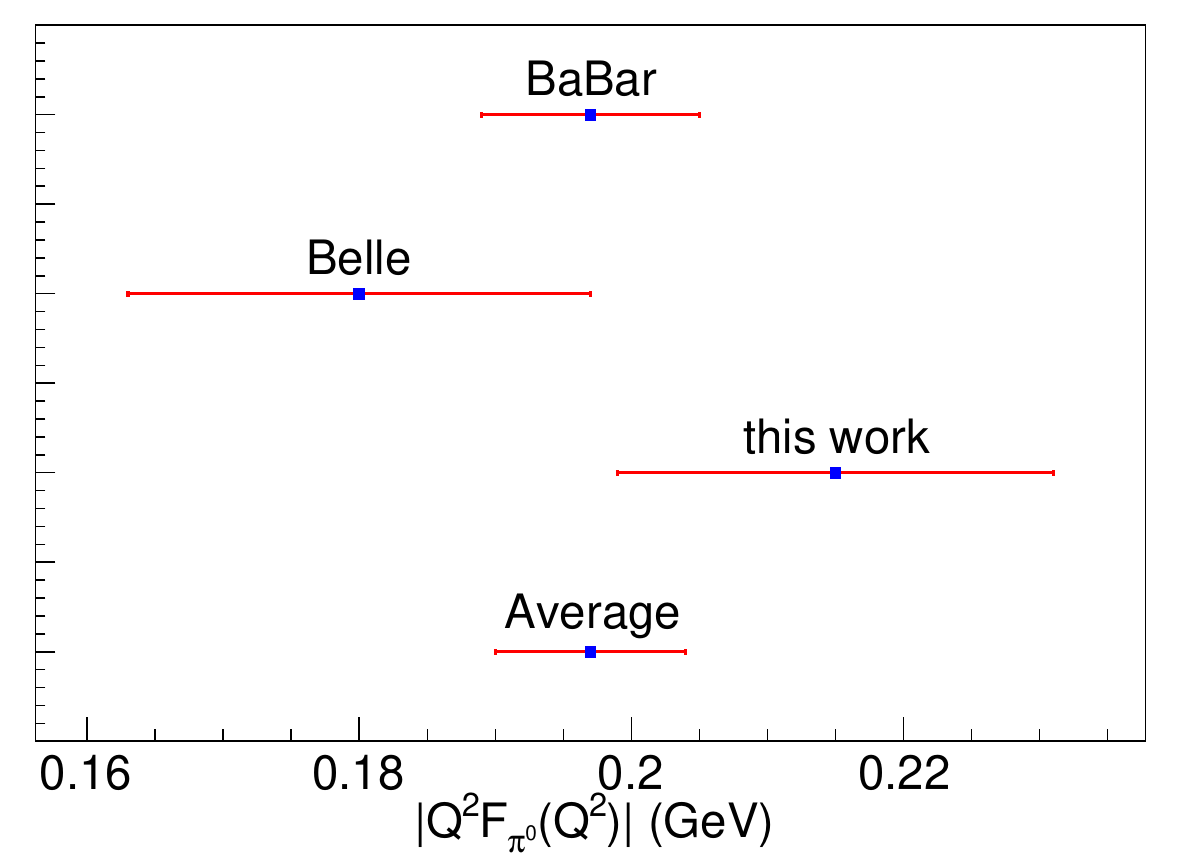}
		\caption{$\left|Q^{2}F_{\pi^{0}}(Q^{2})\right|$ from different measurements at $\left|Q^{2}\right|=m^{2}(\psi(2S))=3.686^{2}~\rm{GeV}^{2}$.}       
		\label{fig:ff}
\end{figure}

BaBar has suggested a parametrization of the form factor with the form
\begin{equation}
    \left|Q^2F_{\pi^{0}}(Q^2)\right|=A\left(\frac{Q^2}{10{\rm GeV}^{2}}\right)^{b},
\end{equation}
where $A$ and $b$ are free parameters~\cite{BaBar}. 
To compare our results with those of BaBar and Belle experiments, the form factors at $\left|Q^{2}\right|=3.686^{2}~\rm{GeV}^{2}$ from BaBar and Belle under this parametrization are calculated as $\left|Q^{2}F_{\pi^{0}}(Q^{2})\right|=0.179\pm0.017$ and $0.197\pm0.008~\rm{GeV}$, respectively~\cite{belle}. Belle has presented another parameterization defined as~\cite{belle}
\begin{equation}
    \left|Q^2F_{\pi^{0}}(Q^2)\right|=\frac{BQ^2}{Q^{2}+C},
\end{equation}
where $B$ and $C$ are free parameters, and 
corresponding value $\left|Q^{2}F_{\pi^{0}}(Q^{2})\right|=0.180\pm0.017~\rm{GeV}$ at $Q^{2}=3.686^{2}~\rm{GeV}^{2}$. Comparison of these results for $\left|Q^{2}F_{\pi^{0}}(Q^{2})\right|$ between this work and other experiments is shown in Fig.~\ref{fig:ff}, where the differences between our result with Belle and BaBar are $1.5\sigma$ and $1.0\sigma$, respectively. The average of the three results are $\left|Q^{2}F_{\pi^{0}}(Q^{2})\right|=0.197\pm0.007$ GeV when $\left|Q^{2}\right|=m^{2}(\psi(2S))=3.686^{2}~\rm{GeV}^{2}$.
\input{acknowledgement_2024-03-31}

\bibliographystyle{apsrev4-2}
\bibliography{myref}



\end{document}

%% file: authorlist_2024-03-31.tex
M.~Ablikim$^{1}$, M.~N.~Achasov$^{4,c}$, P.~Adlarson$^{76}$, O.~Afedulidis$^{3}$, X.~C.~Ai$^{81}$, R.~Aliberti$^{35}$, A.~Amoroso$^{75A,75C}$, Q.~An$^{72,58,a}$, Y.~Bai$^{57}$, O.~Bakina$^{36}$
% ...
\\
\vspace{0.2cm}
(BESIII Collaboration)\\
\vspace{0.2cm} {\it
$^{1}$ Institute of High Energy Physics, Beijing 100049, People's Republic of China\\
% ...
$^{3}$ Bochum  Ruhr-University, D-44780 Bochum, Germany\\
$^{4}$ Budker Institute of Nuclear Physics SB RAS (BINP), Novosibirsk 630090, Russia\\
% ...
$^{35}$ Johannes Gutenberg University of Mainz, Johann-Joachim-Becher-Weg 45, D-55099 Mainz, Germany\\
$^{36}$ Joint Institute for Nuclear Research, 141980 Dubna, Moscow region, Russia\\
% ...
$^{57}$ Southeast University, Nanjing 211100, People's Republic of China\\
$^{58}$ State Key Laboratory of Particle Detection and Electronics, Beijing 100049, Hefei 230026, People's Republic of China\\
% ...
$^{72}$ University of Science and Technology of China, Hefei 230026, People's Republic of China\\
% ...
$^{75}$ University of Turin and INFN, (A)University of Turin, I-10125, Turin, Italy; (B)University of Eastern Piedmont, I-15121, Alessandria, Italy; (C)INFN, I-10125, Turin, Italy\\
$^{76}$ Uppsala University, Box 516, SE-75120 Uppsala, Sweden\\
% ...
$^{81}$ Zhengzhou University, Zhengzhou 450001, People's Republic of China\\

\vspace{0.2cm}
$^{a}$ Deceased\\
% ...
$^{c}$ Also at the Novosibirsk State University, Novosibirsk, 630090, Russia\\
% ...

}

%% file: acknowledgement_2024-03-31.tex
\clearpage
\section*{ACKNOWLEDGMENTS}

The BESIII Collaboration thanks the staff of BEPCII and the IHEP computing center for their strong support. This work is supported in part by National Key R\&D Program of China under Contracts Nos. 2020YFA0406300, 2020YFA0406400, 2023YFA1606000; National Natural Science Foundation of China (NSFC) under Contracts Nos. 11635010, 11735014, 11935015, 11935016, 11935018, 12025502, 12035009, 12035013, 12061131003, 12192260, 12192261, 12192262, 12192263, 12192264, 12192265, 12221005, 12225509, 12235017, 12361141819; the Chinese Academy of Sciences (CAS) Large-Scale Scientific Facility Program; the CAS Center for Excellence in Particle Physics (CCEPP); Joint Large-Scale Scientific Facility Funds of the NSFC and CAS under Contract No. U1832207; 100 Talents Program of CAS; The Institute of Nuclear and Particle Physics (INPAC) and Shanghai Key Laboratory for Particle Physics and Cosmology; German Research Foundation DFG under Contracts Nos. 455635585, FOR5327, GRK 2149; Istituto Nazionale di Fisica Nucleare, Italy; Ministry of Development of Turkey under Contract No. DPT2006K-120470; National Research Foundation of Korea under Contract No. NRF-2022R1A2C1092335; National Science and Technology fund of Mongolia; National Science Research and Innovation Fund (NSRF) via the Program Management Unit for Human Resources \& Institutional Development, Research and Innovation of Thailand under Contract No. B16F640076; Polish National Science Centre under Contract No. 2019/35/O/ST2/02907; The Swedish Research Council; U. S. Department of Energy under Contract No. DE-FG02-05ER41374

